\documentclass[twocolumn,longbibliography]{revtex4-1}

\usepackage{amsmath}
\usepackage{amssymb}
\usepackage{comment}
\usepackage{graphicx}
\usepackage{xcolor}
\usepackage{cancel}
\usepackage[normalem]{ulem}
\usepackage{hyperref}

\usepackage{color}
\usepackage{soul}
\usepackage{url}

\begin{document}

\title{Beyond persuasion: Mobile electoral interfaces in polarized societies}

\author{Nuno Crokidakis$^{1}$}
\thanks{nunocrokidakis@id.uff.br}
\author{Celia Anteneodo$^{2}$}

\affiliation{$^{1}$Instituto de Física, Universidade Federal Fluminense, Niterói/RJ, Brazil \\ 
$^{2}$Departamento de Física, PUC-Rio, Rio de Janeiro/RJ, Brazil
}

\begin{abstract}
\noindent
Many traditional models of electoral dynamics emphasize opinion change, social influence, and persuasion mechanisms. In contrast, contemporary polarized societies often exhibit relatively stable partisan blocs coexisting with an electorally relevant mobile population. We propose a minimal dynamical model in which electoral competition is governed not by direct persuasion between opposing blocs, but by the dynamics and allocation of a mobile electoral interface. The electorate is partitioned into two stable partisan blocs and a mobile fraction, while direct transitions between the polarized blocs are strongly suppressed. Mobile voters are allocated between the competing blocs through a Fermi-like probabilistic rule governed by rejection asymmetries. Analytical calculations show that electoral susceptibility to political shocks is proportional to the stationary size of the mobile population, identifying electoral mobility as the key quantity that controls the macroscopic response of polarized systems.  A complementary microscopic stochastic formulation  based on independent voter choices  yields an exact relation between electoral susceptibility and finite-size fluctuations, and reveals a transient maximum in electoral-choice fluctuations  when a rejection shock reverses which bloc is favored among mobile voters. 
The model also encompasses a continuous range of mobility  levels, from nearly frozen polarization to highly responsive electoral states. These results suggest that electoral changes in strongly polarized societies can arise from the redistribution of mobile voters without requiring large-scale ideological conversion.

\end{abstract}
 

 \maketitle

\section{Introduction}

Political polarization has become a prominent feature of many contemporary democracies~\cite{abramowitz2010disappearing,McCoy2018PolarizationAT,somer2018deja,bertotti2024opinion}. Electoral competition in strongly polarized societies often exhibits a striking combination of political rigidity and electoral uncertainty. A large fraction of the electorate may remain strongly attached to opposing political blocs, while a comparatively smaller group of weakly aligned, independent, or undecided voters retains a higher degree of electoral mobility. Such configurations have been documented in several contemporary democracies, where affective polarization and partisan identity coexist with an electorally relevant fraction of voters whose choices remain comparatively fluid \cite{abramowitz2010disappearing}. In this context, political competition does not necessarily require substantial direct conversion between opposing partisan blocs \cite{mccoy2019toward}: relatively small changes within the more mobile portion of the electorate may become decisive for aggregate electoral outcomes.

Statistical physics' approaches have provided a broad theoretical framework for understanding collective political and social dynamics \cite{Castellano09socialDynamics,martins2002surveys,galam2008review,calvao2015stylized,ramos2015,filho_crokidakis,crokidakis2026corruption}. In particular, sociophysics models have investigated how local interactions, majority rules, contrarian behavior, inflexible agents, and social influence can shape macroscopic opinion states and even electoral outcomes \cite{Galam2004,Galam2012,GalamCheon2020,Galam2025,Galam2026}. Classical voter-like models, bounded-confidence models and related approaches typically describe the evolution of individual opinions through mechanisms such as imitation, social influence, assimilation, persuasion or interaction among agents \cite{Castellano09socialDynamics,martins2002surveys,galam2008review,calvao2015stylized}. These approaches have been extensively used to investigate the emergence of consensus, fragmentation and polarization from microscopic interactions. Recent developments continue to explore these mechanisms in increasingly rich settings, including heterogeneous interaction structures, nonlinear influence, bounded-confidence dynamics, and polarization-dependent updating \cite{Starnini2025,Bernardo2024, Brooks2024,Cyranka2024,AguilarJanita2024}. In particular, recent surveys emphasize that a central question in opinion dynamics remains how individual-level interactions generate or transform macroscopic opinion configurations \cite{Starnini2025,Bernardo2024}.

Electoral models have also recognized that voters need not be equally susceptible to political influence. In particular, distinctions between relatively stable partisan supporters and floating or weakly committed voters have been introduced in quantitative descriptions of elections \cite{Sano2017}. Empirical studies further indicate that increasing political polarization can be accompanied by a reduction in the fraction of voters switching between parties across successive elections \cite{Smidt2017}. These observations suggest that electoral mobility itself may constitute an important macroscopic quantity, distinct from the microscopic evolution of political opinions.

The present work addresses a complementary question to the traditional opinion-dynamics problem. Rather than asking how polarization emerges from interactions among individuals, we consider a situation in which polarization is already established and direct transitions between opposing partisan blocs are strongly inhibited. We then ask how such a system responds to political perturbations when electoral competition is concentrated primarily within a mobile electoral interface. In this perspective, the relevant dynamical question shifts from large-scale ideological conversion to the size and responsiveness of the fraction of voters that remains electorally mobile.

This distinction becomes particularly relevant in political environments characterized by negative partisanship and rejection-driven voting. Electoral decisions need not be determined exclusively by attraction toward a preferred candidate or party; voters may also choose primarily in opposition to an alternative they reject. Recent research has identified negative voting as a relevant component of electoral behavior and has related it to affective polarization and negative partisan identities \cite{Weber2021,GarziaSilva2024,Areal2024}. Particularly interesting for the present framework, recent evidence indicates that political independents are more likely than partisans to hold negatively based electoral preferences and to describe their choices in terms of opposition rather than support \cite{Siev2024}. Negative evaluations of political opponents can also generate substantial behavioral responses in contemporary information environments \cite{Yu2024}. These findings suggest a natural connection between electoral mobility and rejection-driven choice: voters who are less strongly incorporated into partisan blocs may nevertheless make consequential electoral decisions according to the relative rejection of the competing alternatives.

Motivated by these observations, we propose a minimal dynamical model of electoral mobility in a polarized environment. The population is decomposed into comparatively rigid partisan cores and a mobile electoral interface. The first stage of the dynamics determines the size of this interface through competing alignment and detachment processes. Once the mobile fraction reaches its stationary value, its allocation between the two competing blocs is governed by their relative rejection levels. This separation allows us to distinguish the mechanisms determining how many voters remain mobile from those determining how mobile voters are electorally allocated.

The model yields an analytical expression for the electoral susceptibility, which quantifies the response of aggregate vote shares to changes in rejection asymmetry. We show that this susceptibility is proportional to the stationary mobile fraction of the electorate. Consequently, even substantial political shocks may produce weak macroscopic electoral responses when mobility is strongly suppressed. Conversely, when a sizable mobile interface persists, comparatively small changes in rejection can generate significant electoral effects. The resulting framework therefore describes a  smooth crossover  between low- and high-mobility electoral {\it ranges} and provides a minimal mechanism through which strongly polarized systems can remain simultaneously resistant to large-scale partisan conversion and sensitive to  fluctuations in electoral choices within the mobile population.


\section{Model}

\subsection{Population decomposition}

We consider a highly polarized electoral system in which  two major political parties or blocs dominate the political landscape, and where direct transitions between opposing blocs are strongly suppressed.   

The population is divided into three main compartments, with 
 $L(t)$ and $R(t)$ representing the fractions of the population associated with the two polarized political blocs (e.g., left- and right-wing blocs), and $M(t)$ denoting the fraction of the electorate that is not aligned with either of the two main political blocs. 
 
We assume that, within the time window of the electoral period, 
 each compartment contains a rigid nucleus and a dynamical component. 
In particular, the compartment $M(t)$ contains both a rigid nonaligned nucleus, representing individuals who remain permanently outside the two major partisan blocs, and a fluctuating component that participates in the electoral dynamics.

Denoting the rigid nuclei by $L_0$, $R_0$, and $M_0$, the corresponding temporary supporters by $l(t)$ and $r(t)$, respectively, and the mobile interface by $m(t)$, we write

\begin{eqnarray} \nonumber
    L(t) &=& L_0 + l(t), \\ \nonumber
    R(t) &=& R_0 + r(t) \\ \nonumber
    M(t) &=& M_0 + m(t), 
\end{eqnarray}
verifying the normalization condition  $ L(t)+R(t)+M(t)=1$. 
Therefore,

\begin{equation} \label{eq:norm-M}
    m(t) = m_{\rm max} - l(t) - r(t),
\end{equation}
where 
\begin{equation} \label{eq:mmax}
m_{\rm max} =1 - (L_0 + R_0 +M_0)\,
\end{equation}
\noindent
where $m_{\rm max}$ represents the maximum size of the dynamically mobile electoral interface, corresponding to the fraction of the electorate not permanently assigned to the fixed compartments $L_0, R_0$ and $M_0$. It therefore sets the upper bound for the population that can be dynamically redistributed between the competing political blocs.

The rigid fractions $L_0$, $R_0$ and $M_0$ are assumed to remain constant over the electoral period, whereas $l(t)$, $r(t)$ and $m(t)$ evolve dynamically. The alignment and detachment rates are treated as constant parameters in the present analysis, while rejection levels may change over shorter timescales in response to political events. 

In the context of polarized societies, the dynamical component $m(t)$ can be interpreted as a mobile electoral interface separating comparatively stable partisan blocs. Rather than representing a specific ideological group, it consists of voters who are undecided or only weakly attached to either bloc, and whose electoral behavior remains susceptible to short-term political influence, rejection asymmetries, strategic voting, abstention, and campaign-driven fluctuations.

 The term ``interface'' is used here in an effective, non-spatial sense. Unlike interfaces separating ordered domains in conventional statistical-physics models, the mobile electoral interface does not represent a geometrical boundary or a network structure. Instead, it denotes an intermediate population compartment through which transitions between opposing political blocs may occur, rather than through direct conversion between them.  
 Its members can become temporarily aligned with either bloc, and its size, $m(t)$, measures the fraction of the electorate available for such redistribution, rather than a spatial interface width. 
This interpretation allows us to describe electoral mobility without introducing explicit spatial interactions or microscopic domain boundaries.

\subsection{Mobility dynamics}

Based on the assumption that  direct conversions $l \leftrightarrow r$ are neglected, the linear dynamics of the fluctuating components is given by
 
\begin{eqnarray}
  \frac{dl(t)}{dt} &=&  a_L m(t) - d_L \,l(t),  \label{eq:l}\\
    \frac{dr(t)}{dt} &=& a_R  m(t) - d_R \,r(t). \label{eq:r}
\end{eqnarray}
 where the (alignment) parameters $a_L$ and $a_R$ are the rates at which voters belonging to the mobile electoral interface become temporarily aligned with the left and right political blocs, respectively. This process can arise from positive actions that increase the attractiveness of a given bloc.
Conversely, the (detachment) parameters $d_L$ and $d_R$ measure the rate of partisan detachment, through which previously aligned voters return to the mobile interface. Such a process may arise from political disappointment, loss of trust in party leadership, weakening partisan identification, strategic repositioning, or other mechanisms that increase electoral fluidity.
In principle, these coefficients  might  depend on time or on the state of the system, thereby capturing a broader range of electoral processes. For analytical tractability, however, we restrict the present analysis to the homogeneous case.

From Eqs.~(\ref{eq:norm-M})-(\ref{eq:r}), the mobile electoral interface evolves in time as

\begin{eqnarray} \label{eq:M0}
&& \frac{dm(t)}{dt}
    = d_L \,l(t) + d_R \,r(t) - a\, m(t) ,
\end{eqnarray}
where we defined $a = a_L + a_R$  as the total capture rate of the mobile electoral interface by the competing partisan blocs.

\subsection{ Rejection-driven allocation}

Once the mobile fraction is dynamically established, the next step is to determine how it is allocated between the competing blocs. We denote by $\rho_L$ and $\rho_R$ the effective rejection levels associated with blocs $L$ and $R$, respectively. These quantities govern the allocation of the mobile electoral fraction between the competing blocs, and summarize negative evaluations, aversion, scandals, and other political costs associated with each bloc or candidate. Notice that the rejection levels $\rho_L$ and $\rho_R$ should not be confused with the detachment rates $d_L$ and $d_R$ introduced in the mobility dynamics. While the rejection levels describe the current political evaluation of each bloc and determine the allocation of the mobile electorate, the detachment rates quantify the long-term tendency of previously aligned voters to return to the mobile electoral interface. The present formulation treats these quantities as independent effective variables, allowing the contributions of mobility and rejection to be analyzed separately.  A minimal extension incorporating rejection-dependent detachment rates is briefly discussed in Appendix \ref{app:coupling}. 

First, the mobility dynamics determines the size of the mobile electoral interface in a steady state, $m^*$. Second, conditional on this mobile population, a rejection-driven allocation rule determines how these voters are distributed between the competing partisan blocs.

The probability that a mobile voter chooses bloc $L$ is defined as
\begin{equation}  \label{eq:PL}
P_L=\frac{e^{-\beta \rho_L}}{e^{-\beta \rho_L}+e^{-\beta \rho_R}} ,
\end{equation}
varying smoothly between 0 and 1 according to the relative rejection of the competing blocs and the inverse social temperature $\beta$, which controls the sensitivity of mobile voters to rejection. 
For $\beta\to 0$, rejection has little influence on vote choice. For large $\beta$, even small differences in rejection can strongly bias the mobile electorate.

\begin{figure} [b!]
\centering
\includegraphics[width=0.5\textwidth]{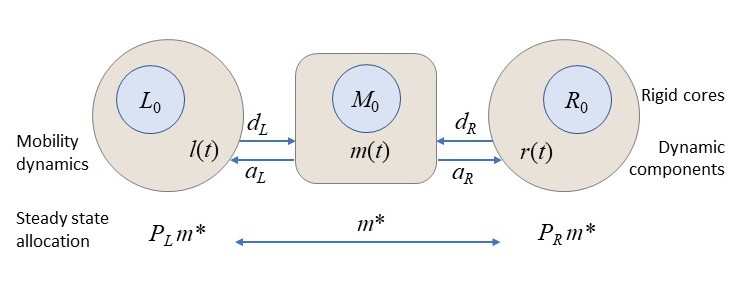}
  \caption{Schematic representation of the model. Each political bloc consists of a rigid partisan core 
  ($L_0$ and $R_0$) and a fluctuating aligned population ($l(t)$ and $r(t)$). The nonaligned compartment is similarly composed of a rigid nonaligned nucleus of size $M_0$ and a mobile electoral interface of size $m(t)$. Alignment and detachment processes determine the size of the mobile interface, with steady state $m^*$, while a rejection-based probabilistic rule allocates the remaining mobile voters between the competing blocs.}  
  \label{fig:scheme}
\end{figure}

Defining
\begin{equation}  \label{eq:delta}
\Delta \rho=\rho_R - \rho_L,
\end{equation}
we can write
\begin{equation}  \label{eq:PLfinal}
P_L=\frac{1}{1+e^{-\beta \Delta\rho}}.
\end{equation}

The complementary probability is $P_R=1 - P_L$. The rejection levels are dimensionless effective quantities, with an arbitrary reference scale. Since only the product $\beta\Delta\rho$ enters the allocation probability, the rejection scale and $\beta$ cannot be independently identified unless a normalization convention is specified. Accordingly, $\beta$ should be interpreted relative to the adopted rejection scale.

Note that the rejection-based allocation rule has the same mathematical structure as the Fermi-Dirac occupation function in quantum statistical mechanics. In this analogy, the rejection difference $\Delta \rho$ plays the role of an effective energy difference,  allowing the model to quantify how asymmetries in candidate rejection influence the redistribution of the mobile electorate.

The model is schematically summarized in Fig. \ref{fig:scheme}


\section{Analytical results}

\subsection{Stationary mobile electorate}

 For clarity, we first consider equal detachment rates, $d_L=d_R=d$, which allows us to isolate the effects
of electoral mobility and rejection asymmetry. The general case of unequal detachment rates is analyzed in Appendix \ref{app:asymmetric}, where we obtain the stationary electoral shares, examine the competition between structural and rejection asymmetries, and determine the transient relaxation rates.

The analytical solution of the mobility dynamics can now be obtained. For the symmetric case, Eq.~\eqref{eq:M0} becomes 
 
\begin{equation} \label{eq:M}
\frac{dm(t)}{dt}=d \, m_{\rm max} - (d + a) m(t),    
\end{equation}
describing the evolution of the mobile electoral interface resulting from the competition between alignment and detachment processes.

Equation~\eqref{eq:M} can be directly integrated, yielding
\begin{equation} \label{eq4}
m(t)=\frac{d }{d + a} \,m_{\rm max}+ C\,e^{-(d + a)t}, 
\end{equation}
\noindent
where $C$ is the integration constant depending on the initial condition. Eq.~\eqref{eq4} leads to the steady state

\begin{equation} \label{eq:steady_M}
m^*=\frac{d }{d + a} \,m_{\rm max}.
\end{equation}
Therefore, the stationary mobile fraction emerges from the dynamical balance between two competing mechanisms: the capture of mobile voters by the partisan blocs (quantified by $a$) and the re-mobilization of previously aligned voters back to the electoral interface (quantified by $d$).  
 
The mobility dynamics relaxes toward the stationary value $m^*$ on the characteristic timescale
\[
\tau=\frac{1}{d+a}.
\]
If this relaxation time is short compared with the duration of the electoral process, the mobile electorate rapidly reaches a quasi-stationary state before significant changes in electoral preferences occur. In this limit, the mobile fraction remains close to $m^*$ throughout the electoral response process, and short-term variations in vote shares arise primarily from the reorganization of the mobile electoral interface rather than from changes in its size or direct large-scale conversion between partisan blocs.

Three mobility ranges can be distinguished: 

(i)  {\bf Low-mobility range:} $m^*\ll m_{\rm max}$.\\
In this  range  ($d\ll a$), electoral mobility is strongly suppressed and the electorate becomes dynamically frozen into stable partisan blocs. Political shocks produce only weak macroscopic changes in vote shares, and electoral competition becomes confined to a narrow mobile interface.

(ii)  {\bf Intermediate mobility:} $m^*=O(m_{\rm max})$.  \\
A sizable mobile electorate coexists with comparatively stable partisan blocs. This  range  may correspond to  polarized electoral systems in which a significant fraction of the electorate  remains dynamically open to strategic voting, rejection asymmetries, or short-term political fluctuations. 
Electoral competition remains significant, but political shocks primarily act through the reorganization of the mobile electoral interface rather than through large-scale ideological conversion.

(iii) {\bf High-mobility  range:} $m^*\approx m_{\rm max}$.\\
A large fraction of the potentially mobile electorate remains dynamically available for electoral reallocation. Electoral competition therefore involves a broad mobile interface, and the aggregate outcome becomes highly responsive to differences in rejection between the competing blocs. In this  range, political shocks affecting the rejection asymmetry can produce substantial macroscopic changes in vote shares.

 It is important to emphasize that the distinction between low, intermediate and high electoral mobility does not correspond to different dynamical phases. In the symmetric model, the normalized stationary mobile fraction, $m^*/m_{\max}=d/(a+d)$, varies smoothly with the ratio $a/d$, without any critical threshold or bifurcation. The three mobility ranges therefore provide a descriptive classification of the competition between alignment and detachment processes, rather than distinct phases separated by dynamical transitions.

The different mobility levels have a clear physical interpretation in terms of the ratio $a/d$, which measures the relative importance of alignment and detachment processes. Low mobility corresponds to alignment processes dominating over detachment ($a\gg d$), while high mobility corresponds to detachment dominating over alignment ($a\ll d$). Intermediate mobility occurs when these processes have comparable rates ($a\sim d$). These distinctions reflect balances between the two processes but do not imply changes in the qualitative structure of the stationary solutions.

The stationary quantities $l^*$ and $r^*$, which represent voters who have become temporarily aligned with the partisan blocs through the mobility dynamics, can be obtained from Eqs.~(\ref{eq:l}) and (\ref{eq:r}), yielding

\begin{equation}
l^*=\frac{a_L}{d+a}\,m_{\rm max},
\end{equation}

and

\begin{equation}
r^*=\frac{a_R}{d+a}\,m_{\rm max}.
\end{equation}

The remaining mobile fraction $m^*$ is assumed to make its electoral choice immediately before the election according to the rejection probabilities $P_L$ and $P_R$.
Consequently, the vote shares  
 of the two blocs, expressed as fractions of the total electorate,  are
 
\begin{equation}  \label{eq:VL}
V_L= L_0 + l^*+P_L \,m^*=  L_0+\frac{a_L +d P_L}{d + a}  \,m_{\rm max}
\end{equation}
and
\begin{equation}  \label{eq:VR}
V_R = R_0 + r^* + P_R \,m^* =  R_0+ \frac{a_R +d P_R}{d + a} \,m_{\rm max}, 
\end{equation}
with $V_L+V_R +M_0 = 1$.

Thus, the comparatively stable partisan blocs contribute directly to the final outcome, while the residual mobile fraction is distributed according to the relative rejection of the competing blocs.

\subsection{Electoral susceptibility}

 A central response relation  of the model is that the macroscopic electoral susceptibility is directly controlled by the stationary size of the mobile electoral interface. Differentiating the stationary vote share $V_L$ with respect to the rejection asymmetry $\Delta\rho$, we obtain

\begin{equation}  \label{eq:CHI}
\chi_e = \frac{\partial V_L}{\partial \Delta\rho} = m^*\beta P_L(1-P_L).
\end{equation}

\begin{figure}[t]
\centering
\includegraphics[width=0.48\textwidth]{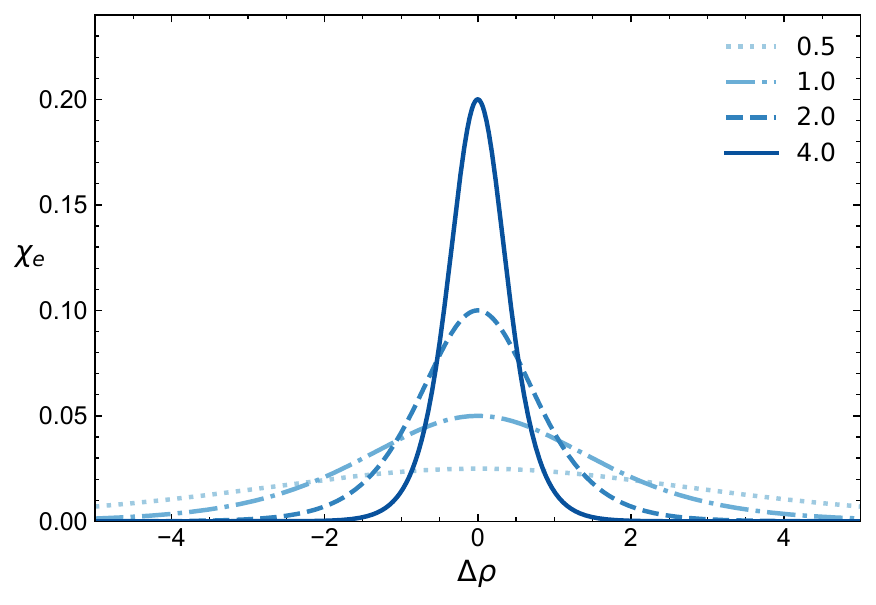}
\caption{Electoral susceptibility $\chi_e$ as a function of the rejection asymmetry $\Delta\rho=\rho_R-\rho_L$, for different values of the inverse social temperature $\beta$ indicated in the legend (increasing from lighter to darker) and fixed $m^*=0.2$. The susceptibility reaches its maximum at $\Delta\rho=0$, corresponding to equally rejected political blocs, and decreases as the magnitude of the rejection asymmetry increases. }
\label{fig:suceptibility}
\end{figure}

Since $P_L(1-P_L)\leq 1/4$,  the susceptibility satisfies
\begin{equation}  \label{eq:CHImax}
\chi_e = \frac{\partial V_L}{\partial \Delta\rho} \leq \frac{\beta m^*}{4}, 
\end{equation}
which provides an upper bound on the electoral response for given values of $m^*$ and $\beta$. Figure \ref{fig:suceptibility} illustrates the dependence of $\chi_e$ on the rejection asymmetry for different values of $\beta$.
Note, however, that,   as follows directly from  Eq.~(\ref{eq:CHI}), all curves   collapse onto a single curve when the rescaled quantities $\chi_e/(\beta m^*)$ and $\beta\Delta\rho$ are used.

Equation~(\ref{eq:CHI}) shows that the electoral susceptibility is proportional to the stationary mobile fraction $m^*$. Consequently,   for a given rejection asymmetry and sensitivity to rejection, the aggregate electoral response  to political shocks  decreases with the stationary mobile fraction. In the limit $m^*\rightarrow0$, the susceptibility vanishes, corresponding to a dynamically frozen polarized electorate.

It is noteworthy that the proportionality between the static electoral susceptibility $\chi_e$ and the
stationary mobile fraction $m^*$, Eq.~\eqref{eq:CHI}, follows directly from the assumed separation between mobile and stable electoral components. Likewise, the upper bound on the susceptibility is a consequence of the Fermi-like allocation rule. These results, therefore, characterize the static response of the minimal model rather than a collective critical phenomenon. As discussed in Sec. IV, the introduction of explicit microscopic stochastic dynamics allows us to extend this description to electoral fluctuations, transient responses, and frequency-dependent susceptibility.

This result provides a simple explanation for the resilience of highly polarized electoral systems to political scandals and other external perturbations. Although such events may increase the rejection associated with one political bloc, their macroscopic electoral impact is ultimately limited by the fraction of voters that remains dynamically mobile.

The implications of this scaling become evident when considering the three mobility ranges introduced previously. Since the electoral susceptibility is proportional to $m^*$, it naturally inherits the behavior associated with each of these ranges.
In the low-mobility range ($m^*\ll m_{\rm max}$), changes in rejection produce only minor variations in aggregate vote shares because electoral competition is restricted to a narrow mobile interface. In the intermediate-mobility range, political shocks have a more substantial impact, acting primarily through the redistribution of the mobile electorate rather than through large-scale ideological conversion. Finally, in the high-mobility range ($m^*\sim m_{\rm max}$), the electorate is most responsive to political influence. However, increasing the size of the rigid blocs reduces the maximal mobility $m_{\rm max}$, thereby limiting the electorate's capacity to respond to political shocks and progressively constraining the range of possible electoral outcomes.

\subsection{Response to political shocks}

If a shock increases the rejection of bloc $L$ by an amount $s$, so that $\rho_L\to \rho_L+s$, then $\Delta \rho\to \Delta \rho-s$. Within the linear-response regime, 
from Eq.~(\ref{eq:CHI}), the electoral variation induced by a rejection shock $s$ is
\begin{equation}
\Delta V_L \simeq -m^*\beta P_L(1-P_L)s.
\end{equation}
Consequently,
\begin{equation}
|\Delta V_L| \leq \frac{\beta m^*}{4}|s|.
\end{equation}

The effect of political shocks is therefore controlled not only by their magnitude $|s|$, but also by the size of the mobile electorate $m^*$ and by the sensitivity of mobile voters to rejection, measured by the parameter $\beta$. In the low-mobility range, the aggregate electoral response remains limited by the small size of the mobile population, even beyond the linear-response regime.
 
Although the short-term electoral response is written in terms of the mobile fraction $m^*$, the parameters $d$ and $a$ remain implicitly present through the stationary solution, Eq.~\eqref{eq:steady_M}. Consequently, the macroscopic impact of political shocks depends indirectly on the balance between re-mobilization processes ($d$) and polarization-driven capture dynamics ($a$).

 The present analysis focuses on short-term electoral responses around a stationary mobile fraction $m^*$, so that electoral shocks act primarily through variations in the allocation probability $P_L$. However, the general relation 
 in Eq.~\eqref{eq:VL} implies that electoral outcomes may also change through variations in
composition of the electorate, including the size of the mobile population and the temporarily aligned fractions. To first order in these variations, we obtain, 
\begin{equation} \label{eq:Delta_V_L}
\Delta V_L  \simeq \Delta l^*  + P_L\,\Delta m^*+m^*\,\Delta P_L ~.
\end{equation}

The three terms in Eq.~\eqref{eq:Delta_V_L} represent distinct contributions to electoral reorganization. The first accounts for changes in the temporarily aligned population associated with bloc $L$, while the second describes the effect of changes in the size of the mobile electorate. The third term represents the redistribution of electoral preferences within the mobile population, driven by variations in rejection asymmetry.

These contributions can be grouped into two sources of electoral reorganization: changes in the composition of the electorate (first two terms) and changes in the allocation of mobile voters between the two political alternatives (third term). Importantly, changes in the size of the mobile electorate generally involve changes in the temporarily aligned populations, so that $\Delta l^*$ cannot be neglected without additional assumptions.

Within the uncoupled formulation, changes in rejection levels affect the allocation probability $P_L$ without modifying the stationary mobile population, whereas changes in alignment or detachment rates alter the stationary composition of the electorate. In the coupled extension discussed in Appendix B, changes in rejection levels can affect both mechanisms simultaneously.

\section{Microscopic stochastic dynamics and electoral fluctuations}

The rejection-driven allocation rule introduced in Sec. II.C specifies the stationary probability that a mobile voter selects one of the two competing blocs. We now provide a microscopic stochastic realization of this rule and
investigate its implications for finite-size electoral fluctuations and
time-dependent responses. The construction is deliberately minimal: individual
mobile voters reconsider their electoral choices independently, while the
rejection levels enter as effective costs of choosing each bloc. In keeping
with the quasi-stationary approximation used in Sec. III, we condition the
stochastic choice dynamics on a fixed mobile population of size $N_m=Nm^*$, where $N$ is the total number of voters. Alignment and detachment processes determine $m^*$ through the mobility equations, but are not explicitly simulated in the stochastic process developed below. Throughout this section, fluctuations refer to the stochastic allocation of electoral choices within the mobile population, rather than fluctuations in the size of the mobile compartment itself. The latter would require a stochastic description of the alignment and detachment processes, which is beyond the scope of the present formulation.

\subsection{Microscopic choice dynamics}

Consider $N_m$ mobile voters, each of whom occupies one of two electoral
choice compartments, $L$ or $R$. Let $\rho_L$ and $\rho_R$ denote the effective
rejection costs associated with these alternatives, and define the rejection
asymmetry $\Delta\rho=\rho_R-\rho_L$ as in Eq.~(7). 
We assign the following
continuous-time transition rates to an individual mobile voter: 
\begin{eqnarray} \label{eq:micro_RL}
w_{R\to L} & = & \gamma\,\frac{1}{1+e^{-\beta\Delta\rho}} ~,  \\ \label{eq:micro_LR}
w_{L\to R} & = & \gamma\,\frac{1}{1+e^{\beta\Delta\rho}} ~,
\end{eqnarray}
\noindent
where $\gamma>0$ sets the rate of electoral reconsideration. These are
heat-bath-type transition rates satisfying
\begin{equation}\label{eq:micro_ratio}
\frac{w_{R\to L}}{w_{L\to R}}=e^{\beta\Delta\rho}=e^{-\beta(\rho_L-\rho_R)} ~.
\end{equation}
Thus, the alternative with the larger effective rejection cost is less likely to be
occupied in the stationary state. Although rejection levels are not physical energies, the statistical-mechanical analogy provides a convenient
parameterization of the relative choice rates. 
 
From the transition rates, the stationary probability that a mobile voter chooses bloc $L$ is given by
\begin{equation} \label{eq:micro_peq}
p_{\rm eq}(\Delta\rho)=\frac{1}{1+e^{-\beta\Delta\rho}} ~,
\end{equation}
\noindent
which coincides with the rejection-driven allocation probability $P_L$ introduced in Eq.~\eqref{eq:PLfinal}. 

The transition rates can be written as 
$w_{R\to L}=\gamma p_{\rm eq}$ and
$w_{L\to R}=\gamma(1-p_{\rm eq})$. Denoting by $p(t)$  the probability that a given
mobile voter chooses $L$, its evolution is governed by
\begin{align} \label{eq:micro_single}
 \frac{dp}{dt}
 &=w_{R\to L}[1-p(t)]-w_{L\to R}p(t)\nonumber\\
 &=\gamma[p_{\rm eq}(\Delta\rho)-p(t)] ~.
\end{align}
For constant rejection levels, the stationary solution is therefore
$p^*=p_{\rm eq}(\Delta\rho)=P_L$, recovering the Fermi-like allocation rule of Eq.~\eqref{eq:PLfinal}. Relaxation towards this stationary state is exponential, with characteristic time $\tau_r=\gamma^{-1}$. Importantly,
detailed balance determines the ratio of transition rates but does
not uniquely fix their absolute timescale. The heat-bath prescription thus represents a
specific minimal dynamical choice. 

We now consider the collective dynamics of $N_m$ mobile voters, assuming that they transition independently, with rates that do not depend on the choices of other voters. Let $n\in\{0,1,\ldots,N_m\}$ denote the number of mobile voters currently choosing $L$. The total rates for transitions $n\to n+1$ and $n\to n-1$ are, respectively, 
\begin{eqnarray} \label{eq:micro_birth}
T_n^+ & = & (N_m-n)\gamma p_{\rm eq} ~, \\ \label{eq:micro_death}
T_n^- & = & n\gamma(1-p_{\rm eq}) ~.
\end{eqnarray}

Denoting by $\mathcal P_n(t)$ the probability of finding $n$ such voters at time $t$, the associated birth-death master equation reads 
\begin{align} \label{eq:micro_master}
 \frac{d\mathcal P_n}{dt}={}&
 T_{n-1}^+\mathcal P_{n-1}
 +T_{n+1}^-\mathcal P_{n+1}\nonumber\\
 &-(T_n^++T_n^-)\mathcal P_n ~,
\end{align}
with vanishing transition rates at the boundaries of the allowed interval.

\subsection{Stationary fluctuations and response}

For time-independent rejection levels, detailed balance at the population 
level gives
\begin{equation} \label{eq:micro_DB}
T_n^+\mathcal P_n^*=T_{n+1}^-\mathcal P_{n+1}^* ~, 
\end{equation}
and hence 
\begin{equation} \label{eq:micro_recurrence}
\frac{\mathcal P_{n+1}^*}{\mathcal P_n^*}
 =\frac{N_m-n}{n+1}\frac{P_L}{1-P_L} ~.
\end{equation}
Successive application of this recurrence yields 
\begin{equation} \label{eq:micro_unnorm}
\mathcal P_n^*=\mathcal P_0^*\binom{N_m}{n}
\left(\frac{P_L}{1-P_L}\right)^n ~.
\end{equation}
Normalization, $\sum_{n=0}^{N_m}\mathcal P_n^*=1$, fixes $\mathcal P_0^*=(1-P_L)^{N_m}$, resulting in the binomial distribution 
\begin{equation} \label{eq:micro_binomial}
\mathcal P_n^*=\binom{N_m}{n}P_L^n(1-P_L)^{N_m-n} ~.
\end{equation}
The corresponding stationary mean and variance are 
\begin{eqnarray} \label{eq:micro_mean}
\langle n\rangle^* & = & N_mP_L ~,\\ \label{eq:micro_var_n}
\operatorname{Var}(n)^* & = & N_mP_L(1-P_L) ~.
\end{eqnarray}

Defining $x=n/N$ as the fraction of the total electorate consisting of mobile voters who currently choose $L$ (rather than the fraction $n/N_m$ within the mobile population), we obtain
\begin{eqnarray} \label{eq:micro_mean_x}
\langle x\rangle^* & = & m^*P_L ~, \\ \label{eq:micro_var_x}
\operatorname{Var}(x)^* & = & \frac{m^*}{N}P_L(1-P_L) ~.
\end{eqnarray}
Consequently, $V_L=L_0+l^*+\langle x\rangle^*$ reproduces Eq.~\eqref{eq:VL}, while
the standard deviation of the mobile contribution to the vote share scales as $N^{-1/2}$ for fixed $m^*$ and $P_L$. The deterministic allocation rule is thus recovered in the large-population limit of this independent-voter process. 

The stationary distribution also establishes a relation between electoral response and spontaneous fluctuations. Differentiating the mean mobile contribution with respect to the rejection asymmetry, at fixed $m^*$, gives 
\begin{equation} \label{eq:micro_chi}
\frac{\partial\langle x\rangle^*}{\partial\Delta\rho}
 =\beta m^*P_L(1-P_L)=\chi_e ~.
\end{equation}
Combining this expression with Eq.~\eqref{eq:micro_var_x}, we obtain
\begin{equation} \label{eq:micro_FDR}
\chi_e=\beta N\operatorname{Var}(x)^* ~.
\end{equation}
This equilibrium fluctuation-response identity also follows  directly from Eq.~\eqref{eq:micro_binomial}, which implies 
$\partial\langle n\rangle^*/\partial\Delta\rho
=\beta[\langle n^2\rangle^*-\langle n\rangle^{*2}]$.
Equation \eqref{eq:micro_FDR} thus connects the
stationary electoral susceptibility to finite-size fluctuations in electoral allocation. It is an exact identity for the present independent-choice model with fixed $N_m$, rather than an indication of critical collective fluctuations. In particular, it characterizes fluctuations in the
allocation of mobile voters between the two alternatives, not fluctuations arising from changes in the sizes of the electoral
compartments.

\subsection{Transient response to rejection shocks}

The master equation also allows us to study the response to an abrupt change in rejection asymmetry. For any function $f(n)$, the birth-death process satisfies the exact moment identity 
\begin{align} \label{eq:micro_generator}
 \frac{d\langle f(n)\rangle}{dt}
 ={}&\big\langle[f(n+1)-f(n)]T_n^+\big\rangle\nonumber\\
 &+\big\langle[f(n-1)-f(n)]T_n^-\big\rangle.
\end{align}
 In particular, choosing $f(n)=n$ yields 
\begin{equation}
 \frac{d\langle n\rangle}{dt}
 =\gamma[N_mP_L-\langle n\rangle].
 \label{eq:micro_mean_dynamics}
\end{equation}
For $f(n)=n^2$, we obtain 
\begin{align}
 \frac{d\langle n^2\rangle}{dt}
 ={}&-2\gamma\langle n^2\rangle
 +\gamma(2N_mP_L+1-2P_L)\langle n\rangle\nonumber\\
 &+\gamma N_mP_L.
 \label{eq:micro_second_dynamics}
\end{align}
Using $\operatorname{Var}(n)=\langle n^2\rangle-\langle n\rangle^2$, these expressions give 
\begin{align}
 \frac{d\operatorname{Var}(n)}{dt}
 ={}&-2\gamma\operatorname{Var}(n)\nonumber\\
 &+\gamma(1-2P_L)\langle n\rangle+\gamma N_mP_L.
 \label{eq:micro_var_dynamics}
\end{align}
 Both moment equations are closed, without invoking a mean-field factorization. 

Suppose that the system is initially stationary at rejection asymmetry $\Delta\rho_0$, corresponding to $p_0=P_L(\Delta\rho_0)$. At $t=0$, the rejection asymmetry changes abruptly to $\Delta\rho_1$, corresponding to
$p_1=P_L(\Delta\rho_1)$, while $N_m$ remains unchanged. The mean evolves as 
\begin{equation}
 \langle n(t)\rangle=N_m u(t)
\end{equation} 
where
\begin{equation} \label{eq:micro_u}
u(t)=p_1+(p_0-p_1)e^{-\gamma t}.
\end{equation}
 Since the initial choices are independent and the subsequent transitions remain independent, the full distribution retains its binomial form, 
\begin{equation} \label{eq:micro_transient_binomial}
 \mathcal P_n(t)=\binom{N_m}{n}u(t)^n[1-u(t)]^{N_m-n},
\end{equation}
 so that 
\begin{equation} \label{eq:micro_transient_var}
\operatorname{Var}[n(t)]=N_m u(t)[1-u(t)] ~.
\end{equation}

 Although the mean changes monotonically from $N_mp_0$ to $N_mp_1$, the variance need not be monotonic. In particular, if the rejection shock drives the mobile electorate across the balanced-choice point, i.e., if
$p_0<1/2<p_1$ or $p_1<1/2<p_0$, the variance reaches its maximum when
$u(t_{\max})=1/2$. Solving this condition gives 
\begin{equation} \label{eq:micro_tmax}
t_{\max}=\frac{1}{\gamma}
 \ln\!\left(\frac{p_0-p_1}{1/2-p_1}\right),
\end{equation}
 with $\operatorname{Var}[n(t_{\max})]=N_m/4$. The argument of the logarithm is positive and greater than unity under the stated crossing condition. Thus, a sufficiently strong rejection shock can generate a transient peak in electoral fluctuations even though the average response is monotonic. This peak follows from the binomial statistics of independent choices and should not be interpreted as a dynamical phase transition.

\begin{figure}[t]
\centering
\includegraphics[width=0.48\textwidth]{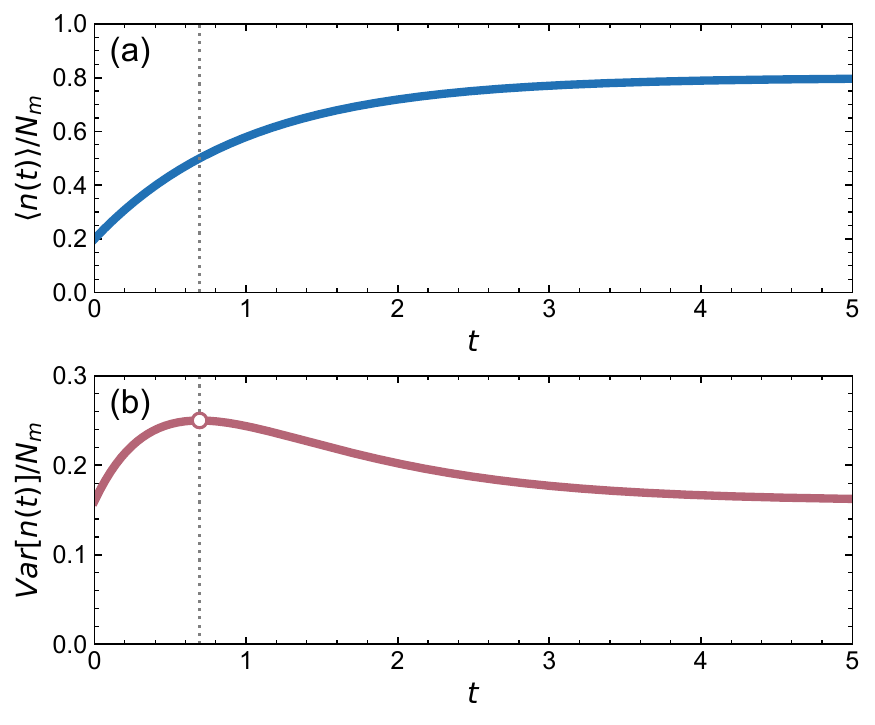}
\caption{Transient response of the mobile electorate following an abrupt change in rejection asymmetry. (a) Time evolution of the mean electoral allocation, normalized by the mobile population size, $\langle n(t)\rangle/N_m$, given by Eq. \eqref{eq:micro_u}. (b) Corresponding normalized variance, $\mathrm{Var}[n(t)]/N_m$, given by Eq. \eqref{eq:micro_transient_var}.  The curves are obtained analytically from the microscopic stochastic dynamics, with $p_0=0.2, p_1=0.8$ and $\gamma=1$. 
While the mean allocation evolves monotonically toward its new stationary value, the variance reaches a transient maximum when the two electoral alternatives are equally represented, 
on average, within the mobile population. 
The dashed vertical line indicates the predicted maximum variance, $t_{\max}=\ln(2)/\gamma \approx 0.693$.}
\label{fig_new}
\end{figure}

Figure 3 illustrates this behavior for a sudden change in rejection asymmetry corresponding to $p_0=0.2$ and $p_1=0.8$, with $\gamma=1$. As shown in Fig. 3(a), the mean
electoral allocation evolves monotonically toward its new stationary value. In contrast, the normalized variance, shown in Fig. 3(b), exhibits a non-monotonic evolution, reaching its maximum at
$t_{\max} \approx 0.693$ given by Eq. \eqref{eq:micro_tmax}. 

 This result highlights a distinction between the average electoral response and its stochastic fluctuations. Even when the mean preference changes monotonically,  fluctuations may be temporarily amplified during the redistribution of the mobile voters between the two political alternatives. Such transient amplification is
absent from the original deterministic formulation and emerges here from the microscopic stochastic dynamics.

\subsection{Frequency-dependent susceptibility}

 Finally, we consider the linear response to a weak, time-dependent rejection asymmetry. From Eq.~\eqref{eq:micro_mean_dynamics}, the mean mobile vote fraction $\bar x(t)=\langle n(t)\rangle/N$ obeys 
\begin{equation} \label{eq:micro_x_dynamic}
\frac{d\bar x}{dt}=\gamma[m^*P_L(\Delta\rho(t))-\bar x(t)].
\end{equation}
 Let $\Delta\rho(t)=\Delta\rho_0+\delta h\,e^{i\omega t}$, where the
physical perturbation is the real part and $|\delta h|$ is small. We write
$\bar x(t)=\bar x^*+\delta x(t)$, with
$\bar x^*=m^*P_L(\Delta\rho_0)$. To first order, 
\begin{align}  \label{eq:micro_linear_logit}
P_L(\Delta\rho(t))\simeq{}&P_L(\Delta\rho_0)\nonumber\\
 &+\beta P_L(\Delta\rho_0)[1-P_L(\Delta\rho_0)]
 \delta h\,e^{i\omega t} ~.
\end{align}
 Substituting into Eq.~\eqref{eq:micro_x_dynamic} and canceling the stationary terms yields 
\begin{equation}
 \frac{d\delta x(t)}{dt}+\gamma\delta x(t)
 =\gamma\chi_e\,\delta h\,e^{i\omega t},
 \label{eq:micro_forced}
\end{equation}
 where $\chi_e$ is the static susceptibility defined in Eq.~\eqref{eq:CHI}. Seeking the
steady periodic response $\delta x(t)=\chi(\omega)\delta h\,e^{i\omega t}$
yields 
\begin{equation} \label{eq:micro_dynamic_chi}
\chi(\omega)=\frac{\gamma\chi_e}{\gamma+i\omega}
 =\frac{\chi_e}{1+i\omega\tau_r} ~,
\end{equation}
 where $\tau_r=\gamma^{-1}$. 
 
 The full linear solution also contains an initial transient proportional to $e^{-\gamma t}$, which decays at long times. For the periodic response, the amplitude and phase are 
\begin{eqnarray} \label{eq:micro_amplitude}
|\chi(\omega)| & = & \frac{\chi_e}{\sqrt{1+(\omega\tau_r)^2}}  \\ \label{eq:micro_phase}
\phi(\omega) & = & -\arctan(\omega\tau_r).
\end{eqnarray}
 Thus, $\chi(\omega\to0)=\chi_e$, while rapid variations in rejection asymmetry 
($\omega\tau_r\gg1$) are attenuated, and the electoral response lags behind the perturbation. The microscopic reconsideration rate therefore introduces a response timescale not specified by the original stationary Fermi rule. 

 The present stochastic formulation recovers the phenomenological allocation rule and static susceptibility as stationary properties of a well-defined continuous-time Markov process, while additionally describing finite-size fluctuations, transient responses to rejection shocks, and a 
 frequency-dependent susceptibility. These results concern the choice dynamics within a quasi-stationary mobile interface. A fully coupled microscopic description
of alignment, detachment and electoral choice would require additional assumptions about how voters entering or leaving the mobile population
acquire or retain their electoral preferences; we leave that extension for future work.


\section{Final remarks}

The model proposed here describes a smooth crossover between mobility ranges, from high-mobility to mobility-suppressed electoral dynamics. In the high-mobility range, a substantial fraction of the electorate remains dynamically available for electoral reallocation, allowing political events to significantly reorganize vote shares. As electoral mobility decreases, however, the mobile electoral interface progressively shrinks, reducing the   macroscopic electoral  response to political shocks. The model therefore identifies the size of the mobile electorate as the key quantity controlling the susceptibility of polarized electoral systems to external political perturbations.
 
The microscopic stochastic formulation,  based on independent voter choices, complements this picture by providing a dynamical foundation for the allocation rule and characterizing finite-size electoral fluctuations. In particular, it reveals that rejection shocks may induce transient amplification of fluctuations even when the mean electoral response is monotonic. The resulting frequency-dependent susceptibility also identifies the timescale of individual electoral reconsideration as an additional factor governing the response to political perturbations.

Recent studies of political polarization suggest that many contemporary democracies are characterized by the coexistence of relatively stable partisan blocs~\cite{somer2018deja,mccoy2019toward} and a substantial fraction of politically mobile, weakly attached, or independent voters~\cite{klar2016independent}. 
 This coexistence is qualitatively consistent with the intermediate-mobility range described by the model. 
 
Within the framework proposed here, electoral competition in polarized settings is driven primarily by the dynamics of a mobile electoral interface rather than by large-scale ideological conversion between strongly committed partisan groups. This interface consists of voters whose political preferences remain comparatively fluid and responsive to current events. Consequently, political shocks, campaign effects, economic conditions, and candidate-specific factors can disproportionately influence electoral outcomes through their impact on this mobile segment of the electorate. Such a scenario is consistent with recent studies emphasizing the coexistence of persistent partisan polarization and a politically mobile electorate in many contemporary democracies~\cite{mccoy2019toward,somer2018deja,klar2016independent}. 

Recent analyses of the Brazilian electorate~\cite{nunes2023biografia}, highlighting the coexistence of relatively rigid partisan blocs with a sizable population of politically mobile or non-aligned voters (``isentões''), are qualitatively consistent with the intermediate-mobility  range  described by the present model.

In such contexts, electoral outcomes may depend less on direct ideological conversion between opposing camps and more on the reorganization of the mobile electoral interface that separates them.

Although the present work is primarily theoretical, the parameters of the model have potential empirical counterparts. Stable partisan fractions and the size of the mobile electorate may, in principle, be estimated from electoral surveys, while rejection levels can be associated with survey-based measures of candidate or party rejection. The dynamical rates $a$ and $d$ would require longitudinal information on transitions into and out of the mobile electorate, ideally obtained from panel surveys tracking the same individuals over time. 
 Given a specified normalization of rejection levels, the sensitivity parameter $\beta$ could in principle be inferred from the dependence of electoral choices on rejection asymmetries.
 Such an empirical calibration lies beyond the scope of the present minimal model, but would provide a natural route toward quantitative applications and predictive tests of the framework.

Although intentionally minimal, the present framework provides a flexible basis for several extensions. Future work may incorporate heterogeneous voter populations, time-dependent alignment and detachment rates, network interactions, multiple political parties, and feedback between electoral shocks and voter mobility. Such extensions  could broaden the applicability of the framework, although their additional complexity would need to be balanced against the simplicity and interpretability of the present formulation.

The present minimal formulation considers two competing political blocs and therefore does not account explicitly for vote fragmentation in multiparty electoral systems. In such systems, rejection of one political alternative may be distributed among several competing parties, potentially producing outcomes that differ substantially from those of an effectively binary competition. A natural extension of the present framework would consist in introducing multiple competing blocs and replacing the binary rejection-driven allocation rule by a multinomial choice rule. Such an extension could allow the model to address vote splitting and other effects characteristic of multiparty elections.

More broadly, the present work suggests a shift in perspective for modeling polarized elections: rather than focusing primarily on large-scale opinion conversion, the electoral response may be understood in terms of the dynamics of a mobile population and fluctuations in its allocation between comparatively stable partisan blocs.

\appendix

\section{Asymmetric detachment rates}\label{app:asymmetric}

 In the main text, we consider equal detachment rates, $d_L=d_R=d$, to isolate the effects of electoral mobility and rejection asymmetry with a minimal number of parameters. Here we relax this assumption and show that the stationary electoral allocation and the transient relaxation remain analytically tractable for $d_L\neq d_R$. 

\subsection{Stationary populations and electoral shares}

 The mobility dynamics is governed by 
\begin{align}
\dot l &= a_L m-d_L l,\\
\dot r &= a_R m-d_R r,
\end{align}
 with $m=m_{\max}-l-r$. For positive detachment rates, stationarity implies 
\begin{equation}
l^*=\frac{a_L}{d_L}m^*,\qquad
r^*=\frac{a_R}{d_R}m^*.
\end{equation}
 Introducing the dimensionless ratios 
\begin{equation}
q_L=\frac{a_L}{d_L},\qquad q_R=\frac{a_R}{d_R},
\end{equation}
 we obtain 
\begin{eqnarray} \label{eq:app_m}
m^* & = & \frac{m_{\max}}{1+q_L+q_R} ~, \\ \label{eq:app_l}
l^* & = & \frac{q_Lm_{\max}}{1+q_L+q_R} ~, \\ \label{eq:app_r}
r^* & = & \frac{q_Rm_{\max}}{1+q_L+q_R} ~.
\end{eqnarray}
 The ratios $q_L$ and $q_R$ quantify the relative strengths of alignment and detachment for each bloc. The equal-detachment case of the main text is recovered by setting $d_L=d_R=d$, so that $q_L+q_R=(a_L+a_R)/d=a/d$. 

 Using the rejection-dependent allocation probability $P_L=[1+\exp(-\beta\Delta\rho)]^{-1}$, the stationary electoral shares become 
\begin{eqnarray} \label{eq:app_VL}
V_L & = & L_0+\frac{m_{\max}(q_L+P_L)}{1+q_L+q_R} ~,\\ \label{eq:app_VR}
V_R & = & R_0+\frac{m_{\max}(q_R+1-P_L)}{1+q_L+q_R} ~.
\end{eqnarray}
 Thus, the composition of the electorate is controlled by $q_L$ and $q_R$, whereas the allocation within the mobile population is controlled by the rejection asymmetry $\Delta\rho=\rho_R-\rho_L$.

\subsection{Competition between structural and rejection asymmetries}

 Subtracting Eqs. \eqref{eq:app_VL} and \eqref{eq:app_VR} yields 
\begin{equation}
V_L-V_R=L_0-R_0+\frac{m_{\max}}{1+q_L+q_R}
\left[(q_L-q_R)+\tanh\!\left(\frac{\beta\Delta\rho}{2}\right)\right].
\label{eq:app_margin}
\end{equation}
 Equation \eqref{eq:app_margin} makes explicit the competition between two distinct sources of electoral asymmetry. The difference $q_L-q_R$ represents the asymmetry associated with alignment and detachment, while the hyperbolic-tangent term describes the rejection-driven allocation of mobile voters. These contributions may reinforce or oppose one another. 

 For a general difference between the stable electoral fractions, the tie condition $V_L=V_R$ can be written as 
\begin{equation}
\tanh\!\left(\frac{\beta\Delta\rho_{\rm tie}}{2}\right)=
(q_R-q_L)-\frac{(L_0-R_0)(1+q_L+q_R)}{m_{\max}}
\equiv B.
\label{eq:app_tie_condition}
\end{equation}
 A finite rejection asymmetry can compensate the structural electoral imbalance only if $|B|<1$, in which case 
\begin{equation}
\Delta\rho_{\rm tie}=\frac{2}{\beta}\operatorname{arctanh}(B).
\label{eq:app_tie}
\end{equation}
 For $|B|>1$, no value of $\Delta\rho$ can produce a tie within this model; for $|B|=1$, a tie is approached only in the limiting case $|\beta\Delta\rho|\to\infty$. In the illustrative case $L_0=R_0$, Eq.~\eqref{eq:app_tie_condition} reduces to $B=q_R-q_L$. Hence, sufficiently strong structural asymmetry cannot be offset by rejection-driven reallocation alone. These are compensation conditions, not critical thresholds or phase transitions. 

 At fixed mobility parameters, the rejection susceptibility retains the same structure as in the symmetric case, 
\begin{equation} \label{eq:app_chi}
\chi_e=\frac{\partial V_L}{\partial\Delta\rho}
=\beta m^*P_L(1-P_L) ~.
\end{equation}
 where $m^*$ is given by Eq.~\eqref{eq:app_m}. The asymmetric rates modify the magnitude of the response through the stationary mobile fraction, but do not alter its logistic dependence on rejection asymmetry. 

\subsection{Transient relaxation}

 Eliminating $m$ from the mobility equations gives the affine linear system 
\begin{equation} \label{eq:app_matrix}
\frac{d}{dt}\begin{pmatrix}l\\r\end{pmatrix}
=\begin{pmatrix}a_Lm_{\max}\\a_Rm_{\max}\end{pmatrix}
-\mathbf M\begin{pmatrix}l\\r\end{pmatrix}
\end{equation}
 where 
\begin{equation}
\mathbf M=\begin{pmatrix}a_L+d_L&a_L\\a_R&a_R+d_R\end{pmatrix}.
\end{equation}
 For $\mathbf y(t)=(l(t)-l^*,r(t)-r^*)^{\mathsf T}$, we have $\dot{\mathbf y}=-\mathbf M\mathbf y$, and therefore 
\begin{equation}
\mathbf y(t)=e^{-\mathbf M t}\mathbf y(0).
\label{eq:app_transient}
\end{equation}
 The two relaxation rates are the eigenvalues of $\mathbf M$, 
\begin{equation} \label{eq:app_eigen1}
\lambda_{\pm}=\frac{S\pm\sqrt{S^2-4D}}{2}
\end{equation}
 where 
\begin{eqnarray} \label{eq:app_eigen2}
S & = & a_L+a_R+d_L+d_R \\ \label{eq:app_eigen3}
D & = & d_Ld_R+a_Ld_R+a_Rd_L.
\end{eqnarray}
 For strictly positive rates, $S>0$, $D>0$, and the eigenvalues are real and positive. Thus, the stationary state is linearly stable, and generic perturbations relax through a superposition of two exponential modes. In particular, for unequal detachment rates, the mobile fraction $m(t)=m_{\max}-l(t)-r(t)$ generally contains contributions from both modes. 

 When $d_L=d_R=d$, the eigenvalues reduce to $\lambda_+=a_L+a_R+d$ and $\lambda_-=d$. The $\lambda_-=d$ eigenmode redistributes the temporarily aligned populations without changing their sum; consequently, the mobile fraction relaxes with the single rate $a_L+a_R+d$ in the symmetric-detachment case. This simplification motivates the use of equal detachment rates in the main text, while Eqs. \eqref{eq:app_m} - \eqref{eq:app_eigen3} establish that the principal stationary conclusions extend to unequal rates.


\section{Rejection-dependent detachment rates}\label{app:coupling}

In the main text, the mobility dynamics and the rejection-driven allocation are treated as separate processes. This assumption allows us to distinguish the mechanisms determining the size of the mobile electorate from those governing its allocation between competing political blocs. Here, we briefly examine a minimal extension in which rejection also affects the detachment rates.

Specifically, we consider
\begin{eqnarray}
d_L & = &d_0+\kappa\rho_L, \\
d_R & = & d_0+\kappa\rho_R,
\end{eqnarray}
where $d_0>0$ is the baseline detachment rate and $\kappa\geq0$ measures the strength of the coupling between rejection and detachment. We assume nonnegative rejection levels, so that both detachment rates remain positive.

Using the general stationary solutions derived in Appendix A, the mobile fraction becomes
\begin{equation}
m^*=\frac{m_{\max}}{1+\dfrac{a_L}{d_0+\kappa\rho_L}+\dfrac{a_R}{d_0+\kappa\rho_R}}.
\end{equation}

The corresponding electoral share of bloc $L$ is
\begin{equation}
V_L=L_0+m^*\left(\frac{a_L}{d_0+\kappa\rho_L}+P_L
\right),
\end{equation}
where $P_L$ is given by Eq. \eqref{eq:PLfinal}.

For a variation in $\rho_L$ at fixed $\rho_R$, differentiation yields
\begin{equation}
\frac{\partial V_L}{\partial\rho_L}
=
-m^*\left[
\beta P_L(1-P_L)
+
\frac{\kappa a_L}{d_L^2}
\frac{1+a_R/d_R-P_L}
{1+a_L/d_L+a_R/d_R}
\right].
\end{equation}

The first contribution corresponds to the rejection-driven redistribution of mobile voters, already present in the original model. The second arises from the dependence of detachment on rejection and accounts for the resulting changes in the composition and size of the mobile electorate. For positive parameters, both contributions are non-positive, so that increasing the rejection of bloc $L$ decreases its stationary electoral share.

This extension illustrates how rejection-dependent detachment introduces an additional channel for electoral response. Nevertheless, the main formulation deliberately treats mobility and rejection as independent mechanisms, allowing their respective contributions to be identified separately. A more general description involving feedback between electoral preferences and mobility dynamics is beyond the scope of the present work.

\section*{Acknowledgments}
 We acknowledge partial financial support from CAPES-Brazil, under Finance Code 001.
 N.C. acknowledges partial financial support from CNPq-Brazil, under Grant No. 308643/2023-2 and No. 406820/2025-2 (Universal).
 C.A. also acknowledges partial financial support from CNPq-Brazil, under Grant No. 308347/2025-0 and No. 406820/2025-2 (Universal), and from FAPERJ-Brazil, under Grant CNE E-26/204.130/2024.

\bibliography{referencias}

@article{calvao2015stylized,
  title={Stylized facts in Brazilian vote distributions},
  author={Calv{\~a}o, Angelo Mondaini and Crokidakis, Nuno and Anteneodo, Celia},
  journal={PLOS ONE},
  volume={10},
  number={9},
  pages={e0137732},
  year={2015},
  publisher={Public Library of Science},
  url={https://journals.plos.org/plosone/article?id=10.1371/journal.pone.0137732}
}

@book{nunes2023biografia,
  author    = {Felipe Nunes and Thomas Traumann},
  title     = {Biografia do Abismo: Como a Polarização Divide Famílias, Desafia Empresas e Compromete o Futuro do Brasil},
  publisher = {HarperCollins Brasil},
  year      = {2023},
  url ={https://harpercollins.com.br/products/biografia-do-abismo-felipe-nunesthomas-traumann?srsltid=AfmBOoqvdmGa8HNSRZMyyKEFAVslPQuaWa1fstpsKLSlepG8c9QarrL0&variant=44543520964774}
}

@book{klar2016independent,
  title     = {Independent Politics: How American Disdain for Parties Leads to Political Inaction},
  author    = {Klar, Samara and Krupnikov, Yanna},
  year      = {2016},
  publisher = {Cambridge University Press},
  address   = {New York},
  isbn      = {9781316500637},
  doi       = {10.1017/CBO9781316471050}
}

@book{abramowitz2010disappearing,
  author    = {Abramowitz, A.},
  title     = {The Disappearing Center: Engaged Citizens, Polarization, and American Democracy},
  publisher = {Yale University Press},
  year      = {2010},
  isbn      = {9780300162882},
  lccn      = {2009033737},
  url       = {https://books.google.com.br/books?id=LVHF3roJPU0C}
}

@article{McCoy2018PolarizationAT,
  author  = {Jennifer McCoy and Tahmina Rahman and Murat Somer},
  title   = {Polarization and the Global Crisis of Democracy: Common Patterns, Dynamics, and Pernicious Consequences for Democratic Polities},
  journal = {American Behavioral Scientist},
  year    = {2018},
  volume  = {62},
  number  = {1},
  pages   = {16--42},
  doi     = {10.1177/0002764218759576}
}

@article{somer2018deja,
  author  = {Somer, Murat and McCoy, Jennifer},
  title   = {D{\'e}j{\`a} vu? Polarization and Endangered Democracies in the 21st Century},
  journal = {American Behavioral Scientist},
  volume  = {62},
  number  = {1},
  pages   = {3--15},
  year    = {2018},
  doi     = {10.1177/0002764218760371}
}

@article{Castellano09socialDynamics,
  title = {Statistical physics of social dynamics},
  author = {Castellano, Claudio and Fortunato, Santo and Loreto, Vittorio},
  journal = {Rev. Mod. Phys.},
  volume = {81},
  issue = {2},
  pages = {591--646},
  numpages = {55},
  year = {2009},
  month = {May},
  publisher = {American Physical Society},
  doi = {10.1103/RevModPhys.81.591}
}

@article{mccoy2019toward,
  title={Toward a Theory of Pernicious Polarization and How It Harms Democracies: Comparative Evidence and Possible Remedies},
  author={McCoy, Jennifer and Somer, Murat},
  journal={The Annals of the American Academy of Political and Social Science},
  volume={681},
  number={1},
  pages={234--271},
  year={2019},
  publisher={Sage Publications Sage CA: Los Angeles, CA},
  url = {https://ssrn.com/abstract=3275846}
}

@article{galam2008review,
  title={Sociophysics: A review of Galam models},
  author={Galam, Serge},
  journal={International Journal of Modern Physics C},
  volume={19},
  number={03},
  pages={409--440},
  year={2008},
  publisher={World Scientific},
  url={https://doi.org/10.1142/S0129183108012297}
}

@article{martins2002surveys,
  author  = {S. G. Alves and N. M. Oliveira Neto and M. L. Martins},
  title   = {Electoral surveys' influence on the voting processes: A cellular automata model},
  journal = {Physica A},
  volume  = {316},
  pages   = {601--614},
  year    = {2002},
  url={https://www.sciencedirect.com/science/article/pii/S0378437102012086?via%3Dihub}
}

@article{Starnini2025,
  title = {Opinion dynamics: Statistical physics and beyond},
  author = {Starnini, Michele and Baumann, Fabian and Galla, Tobias and Garcia, David and I\~niguez, Gerardo and Karsai, M\'arton and Lorenz, Jan and Sznajd-Weron, Katarzyna},
  journal = {Rev. Mod. Phys.},
  volume = {98},
  issue = {3},
  pages = {035004},
  numpages = {72},
  year = {2026},
  month = {Sep},
  publisher = {American Physical Society}
}

@article{Bernardo2024,
  author  = {Bernardo, Carmela and Altafini, Claudio and
             Proskurnikov, Anton V. and Vasca, Francesco},
  title   = {Bounded confidence opinion dynamics: A survey},
  journal = {Automatica},
  volume  = {159},
  pages   = {111302},
  year    = {2024},
  doi     = {10.1016/j.automatica.2023.111302}
}

@article{Brooks2024,
  author  = {Brooks, Heather Z. and Chodrow, Philip S. and Porter, Mason A.},
  title   = {Emergence of Polarization in a Sigmoidal Bounded-Confidence Model of Opinion Dynamics},
  journal = {SIAM Journal on Applied Dynamical Systems},
  volume  = {23},
  number  = {2},
  pages   = {1442--1470},
  year    = {2024},
  doi     = {10.1137/22M1527258}
}

@article{Cyranka2024,
  author  = {Cyranka, Jacek and Mucha, Piotr B.},
  title   = {A polarization opinion model inspired by bounded confidence communications},
  journal = {Journal of the Franklin Institute},
  volume  = {361},
  number  = {16},
  pages   = {107140},
  year    = {2024},
  doi     = {10.1016/j.jfranklin.2024.107140}
}

@article{ramos2015,
author={Marlon Ramos and Jia Shao and Saulo D. S. Reis and Celia Anteneodo and José S. Andrade and Shlomo Havlin and Hernán A. Makse },
  title   = {How does public opinion become extreme?},
  journal = {Scientific Reports},
  volume  = {5},
  pages   = {10032},
  year    = {2015},
  url     = {https://doi.org/10.1038/srep10032}
}

@article{AguilarJanita2024,
  author  = {Aguilar-Janita, Miguel and Blanco-Alonso, Andres and Khalil, Nagi},
  title   = {Polarization-induced stress in the noisy voter model},
  journal = {Physica A: Statistical Mechanics and its Applications},
  volume  = {647},
  pages   = {129840},
  year    = {2024},
  doi     = {10.1016/j.physa.2024.129840}
}

@inproceedings{Sano2017,
  author    = {Sano, Fumiaki and Hisakado, Masato and Mori, Shintaro},
  title     = {Mean Field Voter Model of Election to the House of Representatives in Japan},
  booktitle = {Proceedings of the Asia-Pacific Econophysics Conference 2016
               -- Big Data Analysis and Modeling toward Super Smart Society --},
  series    = {JPS Conference Proceedings},
  volume    = {16},
  pages     = {011016},
  year      = {2017},
  doi       = {10.7566/JPSCP.16.011016}
}

@article{Smidt2017,
  author  = {Smidt, Corwin D.},
  title   = {Polarization and the Decline of the American Floating Voter},
  journal = {American Journal of Political Science},
  volume  = {61},
  number  = {2},
  pages   = {365--381},
  year    = {2017},
  doi     = {10.1111/ajps.12218}
}

@article{Weber2021,
  author  = {Weber, Till},
  title   = {Negative voting and party polarization: A classic tragedy},
  journal = {Electoral Studies},
  volume  = {71},
  pages   = {102335},
  year    = {2021},
  doi     = {10.1016/j.electstud.2021.102335}
}

@book{GarziaSilva2024,
  author    = {Garzia, Diego and Ferreira da Silva, Frederico},
  title     = {Negative Voting in Comparative Perspective},
  publisher = {Palgrave Macmillan},
  address   = {Cham},
  year      = {2024},
  isbn      = {978-3-031-51208-7},
  doi       = {10.1007/978-3-031-51208-7}
}

@article{Areal2024,
  author  = {Areal, Jo{\~a}o},
  title   = {Beyond disdain: Measurement and consequences of negative
             partisanship as a social identity},
  journal = {Electoral Studies},
  volume  = {90},
  pages   = {102831},
  year    = {2024},
  doi     = {10.1016/j.electstud.2024.102831}
}

@article{Siev2024,
  author  = {Siev, Joseph J. and Rovenpor, Daniel R. and Petty, Richard E.},
  title   = {Independents, not partisans, are more likely to hold and express
             electoral preferences based in negativity},
  journal = {Journal of Experimental Social Psychology},
  volume  = {110},
  pages   = {104538},
  year    = {2024},
  doi     = {10.1016/j.jesp.2023.104538}
}

@article{Yu2024,
  author  = {Yu, Xudong and Wojcieszak, Magdalena and Casas, Andreu},
  title   = {Partisanship on Social Media: In-Party Love Among American
             Politicians, Greater Engagement with Out-Party Hate Among
             Ordinary Users},
  journal = {Political Behavior},
  volume  = {46},
  number  = {2},
  pages   = {799--824},
  year    = {2024},
  doi     = {10.1007/s11109-022-09850-x}
}

@article{Galam2004,
  author  = {Galam, Serge},
  title   = {Contrarian deterministic effects on opinion dynamics:
             ``the hung elections scenario''},
  journal = {Physica A: Statistical Mechanics and its Applications},
  volume  = {333},
  pages   = {453--460},
  year    = {2004},
  doi     = {10.1016/j.physa.2003.10.041}
}

@book{Galam2012,
  author    = {Galam, Serge},
  title     = {Sociophysics: A Physicist's Modeling of Psycho-political Phenomena},
  series    = {Understanding Complex Systems},
  publisher = {Springer},
  address   = {New York},
  year      = {2012},
  doi       = {10.1007/978-1-4614-2032-3},
  isbn      = {978-1-4614-2032-3}
}

@article{GalamCheon2020,
  author  = {Galam, Serge and Cheon, Taksu},
  title   = {Asymmetric Contrarians in Opinion Dynamics},
  journal = {Entropy},
  volume  = {22},
  number  = {1},
  pages   = {25},
  year    = {2020},
  doi     = {10.3390/e22010025}
}

@article{Galam2025,
  author  = {Galam, Serge},
  title   = {Democratic Thwarting of Majority Rule in Opinion Dynamics:
             1. Unavowed Prejudices Versus Contrarians},
  journal = {Entropy},
  volume  = {27},
  number  = {3},
  pages   = {306},
  year    = {2025},
  doi     = {10.3390/e27030306}
}

@article{Galam2026,
  author  = {Galam, Serge},
  title   = {Ratio-Dependent Contrarian Activation in Opinion Dynamics},
  journal = {Entropy},
  volume  = {28},
  number  = {4},
  pages   = {443},
  year    = {2026},
  doi     = {10.3390/e28040443}
}

@article{crokidakis2026corruption,
      author={Nuno Crokidakis and Jaime L. C. da C. Filho},
      title={Corruption as a self-sustained collective state in political systems}, 
      journal = {Frontiers in Physics},
      volume = {14},
      pages = {1897453},
      year    = {2026},
      doi     = {10.3389/fphy.2026.1897453},
}

@article{filho_crokidakis,
author = {Filho, Jaime L. C. da C. and Crokidakis, Nuno},
title = {Opinion dynamics under electoral shocks in competitive campaigns},
journal = {International Journal of Modern Physics C},
volume = {38},
number = {6},
pages = {2750054},
year = {2027},
doi = {10.1142/S0129183127500549},
URL = {https://doi.org/10.1142/S0129183127500549},
eprint = {https://doi.org/10.1142/S0129183127500549},
}

@article{bertotti2024opinion,
  title={Opinion dynamics models describing the emergence of polarization phenomena},
  author={Bertotti, Maria Letizia and Menale, Marco},
  journal={Journal of computational social science},
  volume={7},
  number={3},
  pages={2591--2612},
  year={2024},
  publisher={Springer},
  URL = {https://link.springer.com/article/10.1007/s42001-024-00319-x}
}

\clearpage
\newpage

\end{document}